\documentclass[aps,prl,twocolumn,superscriptaddress,showpacs]{revtex4}
\usepackage{graphicx}
\usepackage{latexsym}
\usepackage{amssymb}
\usepackage{amsmath}
\usepackage{amsfonts}
\usepackage{bm}
\usepackage{multirow}
\usepackage{color}
\usepackage{comment}

\DeclareMathAlphabet{\mathbbold}{U}{bbold}{m}{n}

\makeatletter
\newcommand\xleftrightarrow[2][]{%
\ext@arrow 9999{\longleftrightarrowfill@}{#1}{#2}}
\newcommand\longleftrightarrowfill@{%
\arrowfill@\leftarrow\relbar\rightarrow} \makeatother

\begin{document}

\title{Topological Superconductivity in  Twisted Multilayer Graphene}

\author{Cenke Xu}
\affiliation{Department of Physics, University of California,
Santa Barbara, CA 93106, USA}

\author{Leon Balents}

\affiliation{Kavli Institute of Theoretical Physics, Santa
Barbara, CA 93106, USA}

\date{\today}
\begin{abstract}
We study a minimal Hubbard model for electronically driven
superconductivity in a correlated flat mini-band resulting from
the superlattice modulation of a twisted graphene multilayer. The
valley degree of freedom drastically modifies the nature of the
preferred pairing states, favoring spin triplet $d+id$ order with
a valley singlet structure.  We identify two candidates in this
class, which are both {\em topological} superconductors.  These
states support half-vortices carrying half the usual
superconducting flux quantum $hc/(4e)$, and have topologically
protected gapless edge states.
\end{abstract}

\maketitle

Recent experiments\cite{mag01,mag02,chen2018gate} demonstrate
remarkable correlation phenomena in twisted multi-layer graphene
with small twist angles, for which the resulting moir\'e pattern
induces an effective triangular superlattice with a unit cell much
larger than the microscopic one. The superlattice generally
induces mini-bands with a reduced superlattice Brillouin zone.  It
was theoretically predicted that flat mini-bands should exist in
such systems, an effect especially pronounced near ``magic
angles'' in bilayer systems \cite{flat1,flat2,flat3,flat4}. When
the mini-band at the Fermi energy is much narrower than the
effective Coulomb interaction energy per electron, then
correlation effects may be expected. Experiments on
bilayers\cite{mag01} and trilayers\cite{chen2018gate} find evidence
for a correlated Mott insulating state when such a mini-band
contains an integer number of electrons per superlattice unit
cell. Furthermore, gate tuning the charge density away from the
half-filling bilayer moir\'e Mott insulator with 2 electrons per
unit cell led to superconductivity with strong coupling
characteristics\cite{mag02}. Many features are strikingly similar
to those of the cuprate high-T$_c$ materials, for which
superconductivity also occurs in close proximity to a Mott
insulator. This raises the intriguing possibility of graphene
moir\'e superlattices serving as a new platform for unconventional
superconductivity with unprecedented in-situ tunability. The goal
of the current work is to understand the nature of the observed
superconducting phases. We argue that even in the simplest
situation, the valley degree of freedom of graphene leads to
dramatic modifications to the superconductivity: the preferred
states are {\em topological} superconductors with a valley singlet
structure.

Our results are based on the minimal description of a correlated
flat band in terms of a Hubbard model, with a single ``site'' per
unit cell. This is valid when the superlattice period is large,
and when the inter-band mixing may be neglected. For such a flat
band, the (weak) tunneling between nearest-neighbor unit cells
dominates the kinetic energy. The large period suppresses
interactions beyond nearest-neighbor sites. Furthermore, each unit
cell effectively hosts two degenerate orbital wave functions for
electrons, which correspond to the two original valleys at the
Brillouin zone corners, since the large unit cell moir\'e
modulation cannot mix these states due to their large momentum
space separation.   Our starting point is therefore a two-orbital Hubbard
model on the triangular lattice, with in total four flavors of
single-electron states on each site, including both the spin and
orbital degrees of freedom:
\begin{equation}
H = -t \sum_{\langle ij\rangle}\sum_{\alpha=1}^4 \left(
  c^\dagger_{i,\alpha} c^{\vphantom\dagger}_{j,\alpha} + {\rm
    h.c.}\right) + U \sum_j\left( \sum_{\alpha = 1}^4 n_{j,\alpha}\right)^2. \label{h1}
\end{equation}
Eq.~\ref{h1} has an SU(4) symmetry which corresponds to the
rotation between the four flavors of electron states.

\begin{figure}[tbp]
\begin{center}
\includegraphics[width=220pt]{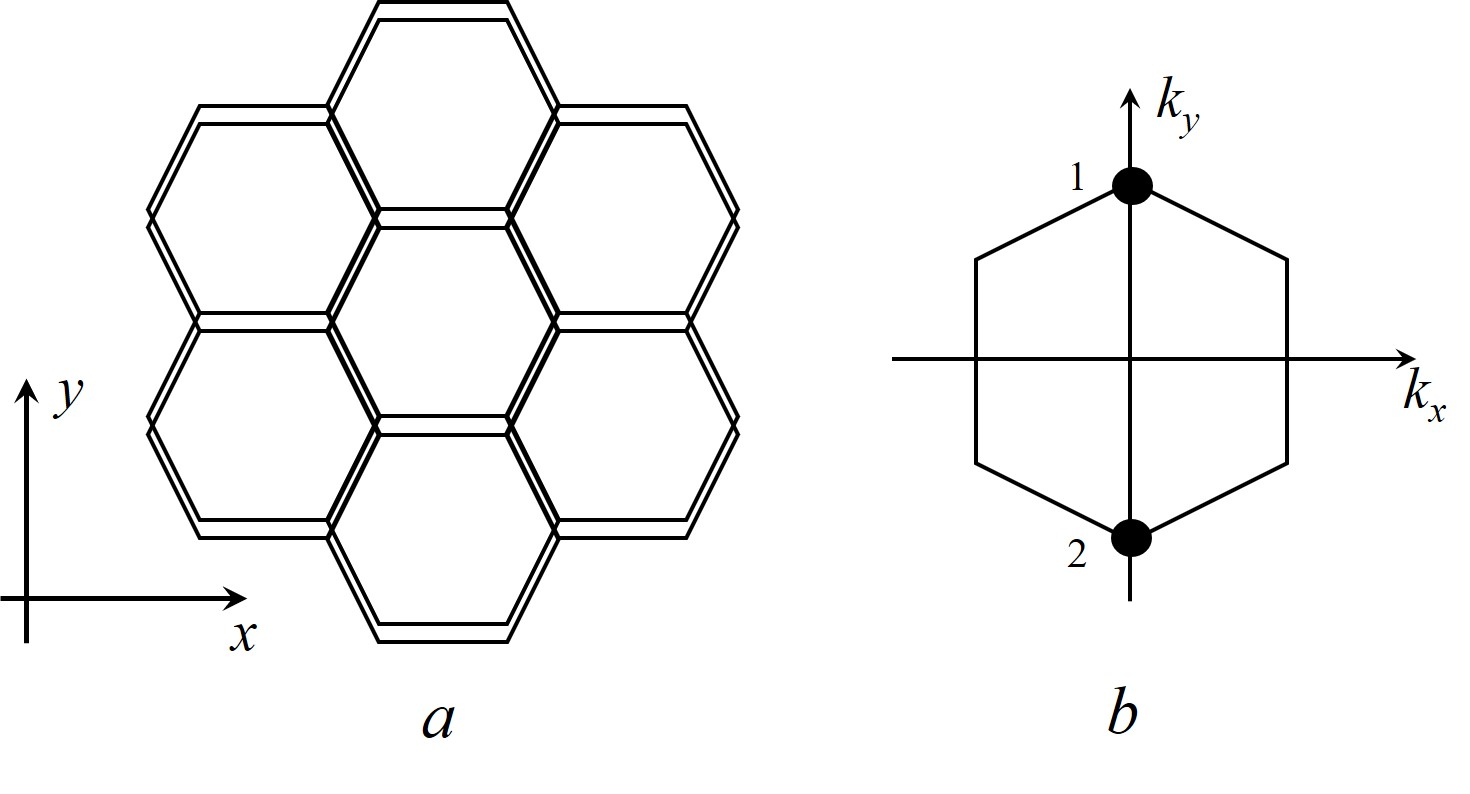}
\caption{ $a$.  The center of a unit cell of a bilayer graphene
moir\'e superlattice, where Wannier states are peaked in the flat
band. In this region there is AA stacking of the two layers, and
coordinates referred to in the text are defined as shown. $b$.
Independent single-particle states are built from momenta near
each of the two valleys shown as dark dots in the microscopic
Brillouin zone corners, and these become the two orbitals in the
Hubbard model of Eq.~\eqref{h1}. } \label{graphene}
\end{center}
\end{figure}

This symmetry is justified as follows.  For the hopping term,
SU(2) spin-rotation invariance requires the hopping to be
spin-independent. Mixing between different orbital states is
prohibited by the large valley separation in momentum space.  The
reality and equality of the hopping amplitudes for the two
different orbitals follows, at least for twisted bilayer graphene
(Fig.~\ref{graphene}), by careful consideration of $2\pi/3$
rotation, reflection $y\rightarrow -y$ (which exchanges the
valleys) and reflection $x\rightarrow -x$. Thus the SU(4) symmetry
of the hopping term in Eq.~\ref{h1} should be an excellent
approximation.  The SU(4) symmetry of the $U$ term follows from
its dependence only on the {\em total} charge of a site, which
physically represents the capacitive energy of a superlattice unit
cell due to ``medium range'' Coulomb interactions, i.e. on scales
large compared to the microscopic lattice spacing but small
compared to the screening length. Corrections to this SU(4)
symmetry arising from short-range interactions do exist and will
be considered later, but are weaker than the dominant SU(4) part
by a factor proportional to $a/a_0$, where $a$ is the superlattice
spacing and $a_0$ is the
microscopic lattice spacing.  

When the number of electrons per site of Eq.~\eqref{h1} is
$\bar{n}=1,2$ or $3$, and when $U/t$ is sufficiently large, the
system becomes a Mott insulator. 
An effective SU(4) Heisenberg model for the Mott insulator can be
derived using the standard perturbation theory based on the
Hubbard model Eq.~\ref{h1}:
\begin{eqnarray}
H_{J} =
J \sum_{\langle ij\rangle}\sum_{a = 1}^{15} \hat{T}^a_i \hat{T}^a_j,
\label{hj}\end{eqnarray}
where $\hat{T}^a_i = c^\dagger_{i,\alpha}
T^a_{\alpha\beta} c^{\vphantom\dagger}_{i,\beta}$,
$c_{i\alpha}^\dagger c_{i\alpha}^{\vphantom\dagger} = \bar{n}$ and $T^a_{\alpha\beta}$ with $a = 1,
\cdots 15$ are fifteen $4\times 4$ Hermitian matrices that form
the fundamental representation of the SU(4) Lie-algebra (we choose
${\rm Tr}\, T^a T^b = 4\delta^{ab}$). The SU(4)
spin model Eq.~\ref{hj} itself is already an interesting subject
to study, and compared with SU(2) spin systems, it is more likely
to support exotic spin liquid ground
states~\cite{wu1,wu2,xuwu,hermele1,savary2015quantum,hermele2,hermele3,zhangsu4,sun,kaulsu4,masakisu4,honeysu4}.
But in this work we will focus on the superconductor phase next to
the Mott insulator after doping.

The Heisenberg interaction in Eq.~\ref{hj} can be rewritten in a different
form (a Fierz identity\cite{savary2017superconductivity}):
\begin{eqnarray}
H_{J} = J \sum_{\langle ij\rangle} \left[-\frac{5}{4}
\left(\vec\Delta_{ij}\right)^\dagger \cdot \vec\Delta_{ij} +
  \frac{3}{4} \left({\sf\Delta}^{-}_{ij}\right)^\dagger \cdot{\sf\Delta}^{-}_{ij} \right] ,
\label{hj2}
\end{eqnarray}
where we defined the 6 component $\vec\Delta_{ij} =
\vec\Delta_{ji}$ and 10 component
${\sf\Delta}^{-}_{ij}=-{\sf\Delta}^{-}_{ji}$ pairing fields
symmetric and anti-symmetric, respectively, in $i\leftrightarrow
j$.  Obviously, the anti-ferromagnetic interactions ($J>0$)
appropriate near half-filling favor condensing the operators
$\vec\Delta_{ij}$, which are ``even parity'' in this sense, and we
henceforth neglect the odd parity channel. In fact,
$\vec{\Delta}_{ij}$ transforms as an SO(6) vector
(SU(4)$\sim$SO(6)), when written in an appropriate basis:
\begin{eqnarray}
\vec{\Delta}_{ij} = c_i^t\left( \sigma^{32}, i\sigma^{02},
\sigma^{12}, i\sigma^{23}, \sigma^{20}, i\sigma^{21} \right) c_j,
\end{eqnarray}
where $\sigma^{ab} = \sigma^a \otimes \sigma^b$, and
$\sigma^0 = \mathbf{1}_{2\times 2}$. The six-component vector
$\Delta^a$ can be decomposed into a three component spin-triplet
and orbital-singlet pairing vector $\bm{\Delta}=\left( \Delta^1, \Delta^2,
\Delta^3 \right)$, and another three component spin-singlet and
orbital-triplet pairing vector $\left( \Delta^4, \Delta^5,
\Delta^6 \right)$. With the SU(4) symmetry of Eq.~\ref{h1} and
Eq.~\ref{hj}, these two sets of three-component vectors are
exactly degenerate.

Upon doping (for instance hole-doping), one can turn on a kinetic
term on Eq.~\ref{hj2}, and the large $U$ limit becomes a t-J model
with a projection that prohibits more than two particles per site.
In an intermediate coupling scenario we can simply add $H_J$ to
the Hamiltonian to represent the effects of antiferromagnetic
fluctuations. Then a standard mean field theory leads to a
condensate of $\vec{\Delta}_{ij}$, i.e. superconductivity.  Due to
Fermi statistics, this even parity pairing is antisymmetric in
SU(4) flavor space, which is the essence of superexchange that
favors antiferromagnetism. Amongst the even parity channels,
$s-$wave pairing is penalized by the large on-site Hubbard $U$
interaction, and we expect $d-$wave pairing to be favored.
Previous studies for SU(2) superconductors on the triangular
lattice found that, to ensure the entire Fermi surface is gapped,
$d_{x^2 - y^2} + i d_{xy}$ pairing is often
favored~\cite{did1,did2,did3,did4,did5,did6,did7}.

\begin{figure}[tbp]
\begin{center}
\includegraphics[width=180pt]{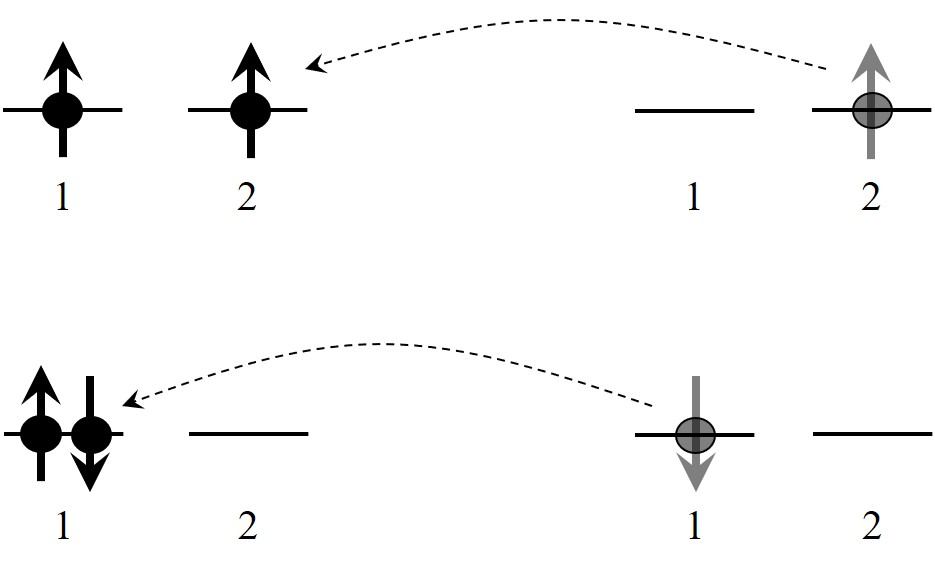}
\caption{The virtual process in second order perturbation theory
in $t/U$.  An electron hops from the singly occupied site on the
right to the one on the left, making it doubly occupied.  The
intermediate state may be (top) a spin triplet and orbital singlet
or (bottom) a spin singlet and orbital triplet. The Hunds
interaction favors the upper situation.} \label{virtual}
\end{center}
\end{figure}

Now let us consider the effects of SU(4) symmetry-breaking
perturbations to the Hubbard model. The dominent effects arise
from interactions, which are analogous to Kanamori terms
multi-orbital Hubbard systems. As is usually the case for
transition metal ions, we assume that the most important of these
is the Hunds coupling
\begin{eqnarray}
H_{h} = - V \sum_j
\left(\bm{S}_{j}\right)^2, \label{hunds}
\end{eqnarray}
where $V> 0$ and $\bm{S}_{j}$ is the total spin on site $j$. The
Hunds coupling $H_h$ is expected to further prefer the
three-component spin-triplet and orbital singlet pairing vector
$\bm\Delta$ over the other three components of the SO(6) vector
$\vec{\Delta}$. To see this, consider two nearest neighbor sites
that are both doped with one hole, $i.e.$ each site is occupied by
one electron, at second order perturbation theory in $t/U$.
Suppose the two electrons form a spin-singlet and orbital triplet
state, then the virtual intermediate state contains one doubly
occupied site which increases the energy relative to two singly
occupied sites by $2U$; while if the two electrons form a
spin-triplet state, then the virtual intermediate state has energy
$2U - 2V$, which is lower than the previous case due to the Hunds
interaction Eq.~\ref{hunds} (Fig.~\ref{virtual}). Thus the Hunds
coupling will select the spin-triplet and orbital-singlet
components from the SO(6) vector $\vec{\Delta}$, and the energy
splitting is at the order of $\sim t^2/(2U - 2V) - t^2/(2U) \sim V
t^2/(2U^2)$. The analysis of the electron-doped case leads to the
same conclusion. Instead of the two site argument, one may
alternatively just consider the modification of the super-exchange
interaction of Eq.~\eqref{hj2} by the $V$ term.  This leads to a
ferromagnetic contribution purely in the spin sector $\sim - J
(V/U) \sum_{\langle ij\rangle} \bm{S}_i \cdot \bm{S}_j$, which by
a similar Fierz identity favors triplet pairing.

It is noteworthy that the valley degree of freedom allows the
formation of an even parity (d-wave) spin triplet state, which is
impossible due to Fermi statistics for a single orbital model.
Here it occurs because the orbital singlet is anti-symmetric.
However, in our discussion we defined the parity and angular
momentum of the pair with respect to the two-orbital Hubbard
model. Microscopically, parity also exchanges the two valleys, so
in terms of the large microscopic Brillouin zone, the even parity
d-wave state becomes an odd-parity f-wave one. We stick with the
former convention for concreteness.

Knowing that the system favors spin-triplet $d+id$ or $d\pm id$
pairing, most generally we can write the spin triplet Cooper pair
matrix in the BdG Hamiltonian as
\begin{eqnarray}
\bm\Delta_k = \left( u_k
\bm{\Phi}_1 + v_k\bm{\Phi}_2 \right) \cdot i \sigma^2
\bm{\sigma} \otimes \sigma^2, \label{pairing}
\end{eqnarray}
where
$u_k = \cos k_x - \cos \frac{k_x}{2} \cos \frac{\sqrt{3} k_y}{2}$ and
$v_k = \sqrt{3} \sin \frac{k_x}{2} \sin \frac{\sqrt{3} k_y}{2}$ are
the periodic superlattice analogs of the $k_x^2-k_y^2$ and $2k_x k_y$
pairing functions, respectively.  Here
$\bm{\Phi}_1$ and $\bm{\Phi}_2$ are both {\it complex} SO(3)
spin vectors. To minimize the energy and maximize the pairing gap
on the Fermi surface, there are two candidate states which are
degenerate at the mean field level:
\begin{eqnarray}
\mathrm{A} &:& \bm{\Phi}_2 = i \bm{\Phi}_1 = i \bm{\phi} e^{i\theta}, \nonumber \\
\mathrm{B} &:& \bm{\Phi}_1 = \bm{\phi}_1 e^{i\theta}, \ \
\bm{\Phi}_2 = \bm{\phi}_2 e^{i\theta}.
\end{eqnarray}
Here $\bm{\phi}$, $\bm{\phi}_1$, $\bm{\phi}_2$ are all
three-component real vectors under spin SO(3) rotation, and
$\bm{\phi}_1 \cdot \bm{\phi}_2=0$. Other types of spin-triplet
superconductors, for example $\bm\Delta_k \sim (u_k+iv_k)
(\bm{\phi}_1 + i \bm{\phi}_2)$, with real vectors $\bm{\phi}_1
\cdot \bm{\phi}_2=0$, do not have a uniform maximal gap on the
Fermi surface, and are thus less favorable within mean field
theory than types A and B.

Type A and B states are degenerate within the standard BCS mean
field theory. This is apparent from comparing for example the type
A state with $\bm{\phi} \sim (0,1,0)$ and the type B state
$\bm{\phi}_1 \sim (1,0,0)$ and $\bm{\phi}_2 \sim (0,1,0)$. In the
former, both spin up and down electrons experience $d+id$ pairing,
while in the latter, the pair field for up spin electrons is
$d+id$, and the pair field for down electrons is $d-id$. The gap
magnitudes are everywhere identical in the two cases, and hence
they have the same mean field energy. This is the consequence of
an artificial symmetry in the mean field formalism: a reflection
symmetry $k_x \rightarrow - k_x$ on spin down electrons only,
which interchanges the two types of pairings. Taking the most
general form of the pairing order parameter $\bm{\Phi}_{\vec{k}} =
\left( u_k \bm{\Phi}_1 + v_k \bm{\Phi}_2 \right)$ with complex
vectors $\bm{\Phi}_1$, $\bm{\Phi}_2$, the BCS mean field theory
generates a Landau-Ginzburg free energy
\begin{eqnarray}
F =
\sum_{\vec{k}} r |\bm{\Phi}_{\vec{k}}|^2 +
g(|\bm{\Phi}_{\vec{k}}|^2)^2 - c |\bm{\Phi}_{\vec{k}} \cdot
\bm{\Phi}_{\vec{k}}|^2.
\end{eqnarray}
The last term maintains the degeneracy between type A and type B
pairing, but disfavors other types of pairings.  In general,
effects beyond the BCS treatment will generate additional terms in
the Landau-Ginzburg free energy and lift the degeneracy between
type A and B. We will not attempt to resolve which state is
favored here, but simply discuss the properties of both candidate
states.

Consider time-reversal symmetry, which flips spin {\em and}
exchanges the two valleys, hence $c \rightarrow \sigma^{21} c$. It
also induces complex conjugation, so it acts on the order
parameter $\bm{\Phi}_{\vec{k}}$ as $\mathcal{T}:
\bm{\Phi}_{\vec{k}} \rightarrow \bm{\Phi}^\ast_{-\vec{k}}$. Thus
type A pairing breaks time-reversal symmetry because $u_k+iv_k
\rightarrow u_k - iv_k$ under complex conjugation, while type B
pairing is time-reversal invariant.

Now consider the topology of the order parameter. Within a single
time-reversal sector, the type A state has the ground state
manifold $[S^2 \times S^1]/Z_2$. Here $S^2$ corresponds to the
configuration of the spin SO(3) vector $\bm{\phi}$, $S^1$
corresponds to the configuration of $e^{i\theta}$.  The full order
parameter is invariant under a $Z_2$ transformation
\begin{eqnarray}
\bm{\phi}
\rightarrow - \bm{\phi}, \qquad \theta \rightarrow \theta + \pi.
\label{z2}
\end{eqnarray}
Due to this, type A pairing supports a half-vortex, analogous to
that in the polar state of spin-1 bosons in cold atom
systems~\cite{spinor,spinor1,spinor2}. After tracing along a full
circle around the half-vortex core, both $\bm{\phi}$ and
$e^{i\theta}$ acquire a minus sign (while $\bm\Delta$ remains
single valued). The half-vortex carries a quantized magnetic flux
\begin{eqnarray}
\Phi_0 = \frac{hc}{4e},
\end{eqnarray}
which is half of the magnetic flux quantum of ordinary
superconductors. Moreover, as was discussed in
Ref.~\cite{spinor2}, in this purely two dimensional
superconductor, the Mermin-Wagner theorem dictates that SO(3)
vector $\bm{\phi}$ is disordered at infinitesimal temperature due
to thermal fluctuations. Hence the system no longer has long-range
or even quasi-long-range order of $\bm{\Delta}$. Instead, what
persists are power-law correlations of a spin-singlet charge-$4e$
order parameter $\bm{\Delta}\cdot \bm{\Delta} \sim e^{2i\theta}$.
The Kosterlitz-Thouless transition out of this algebraic
charge-$4e$ superconducting phase is driven by unbinding of
half-vortices, which leads to a universal superconductor phase
stiffness jump $8T_c/\pi$ at the transition~\cite{spinor2}.

While type B pairing does not break time-reversal symmetry, it has
similar finite temperature behavior. The vectors $\bm{\phi}_1$ and
$\bm{\phi}_2$ are disordered immediately by infinitesimal
temperature, and the system effectively becomes an algebraic
charge $4e$ superconductor with a half-vortex that carries
$hc/(4e)$ magnetic flux.

Both type A and type B superconductors are topological, in the
sense that they both have gapless edge states at their boundary.
In the type A superconductor, the boundary has eight channels of
chiral Majorana fermions, which in the ideal case leads to a
thermal Hall conductance
\begin{eqnarray} \kappa_{xy} = \frac{4
\pi^2 k_B^2 T}{3 h}.
\end{eqnarray}
The edge states of the type-A
superconductor are stable against any disorder and interaction
because they are chiral and hence no backscattering can occur. In
the type A superconductor, because the spin symmetry is
spontaneously broken down to U(1), one spin component is still
conserved: for $\bm{\phi} \sim (0,0,1)$, this is $S^z$. In this
case, it is convenient to introduce a new basis of fermion, for
orbital (valley) $1$, define $\psi_{\alpha,1} = c_{\alpha,1}$; for
orbital $2$, define $\psi_{\alpha,2} = \sigma^2_{\alpha\beta}
c^\dagger_{\beta,2}$, $\alpha,\beta =  \uparrow, \downarrow$. Then
the entire mean field Bogoliubov-de Gennes Hamiltonian for
quasiparticles reads
\begin{eqnarray}
\hat{H} &=& \sum_{\vec{k}}
\psi^\dagger_{\vec{k}} \mathcal{H}(\vec{k}) \psi_{\vec{k}}, \nonumber \\
\mathcal{H}(\vec{k}) &=& \epsilon_k \sigma^{03} +
\Delta \left(u_k \sigma^{31} +  v_k
\sigma^{32}\right). \label{eq:1}
\end{eqnarray}
In this basis, spin-up and spin-down fermions $\psi_{\uparrow}$,
$\psi_{\downarrow}$ both have Hall conductivity $\sigma_{xy} = 2$,
which is visible in Eq.~\eqref{eq:1}
because the pair field acts
in the orbital space (second index $\nu$ of $\sigma^{\mu\nu}$) as a
vector in the $1-2$ plane which winds twice around the origin in
momentum space.
Hence the eight channels of chiral Majorana fermion edge states
can be reorganized into two channels of chiral edge states each
for $\psi_{\uparrow}$ and $\psi_{\downarrow}$. Thus the system
also has a ``spin quantum Hall" conductance $\sigma^s_H = 4$:
namely, if we couple the system to a ``spin gauge field"
$A^s_\mu$, and spin-up, spin-down electrons carry gauge charge
$\pm 1$ under the spin gauge field $A^s_\mu$, then after
integrating out all the fermions, the system generates a level-4
Chern-Simons term for the background spin gauge field:$
\mathcal{L}_{cs} = \frac{4}{4\pi} \epsilon_{\mu\nu\rho}
A^s_\mu\partial_\nu A^s_\rho$.

In the type B superconductor, the boundary has four channels of
counter propagating non-chiral Majorana fermions, and there is no
thermal Hall effect. The stability of the edge states of type-B
superconductor deserves a bit more discussion. Let us again take
$\bm{\phi}_1 \sim (1,0,0)$, and $\bm{\phi}_2 \sim (0,1,0)$, then
this superconductor can be simply viewed as spin-up electrons and
spin-down electrons forming $d+id$ and $d-id$ topological
superconductors separately, and its edge state Hamiltonian reads
\begin{eqnarray}
H_{1d} = \int dx \ \sum_{\alpha = 1}^4 \ \chi_{L,\alpha}
i\partial_x \chi_{L,\alpha} - \chi_{R,\alpha} i\partial_x
\chi_{R,\alpha}. \label{1d}
\end{eqnarray}

The order of $\bm{\phi}_1$
and $\bm{\phi}_2$ fully breaks SO(3) spin symmetry, while the
$Z_2$ symmetry in Eq.~\ref{z2} (a product of $\pi-$rotation in the
spin and charge sectors) is preserved. The $Z_2$ symmetry
acts on the quasiparticles of the superconductor as a fermion
parity for the right-moving fermion $\chi_{R,\alpha}$ only:
$\chi_{L,\alpha} \rightarrow \chi_{L,\alpha}$, $\chi_{R,\alpha}
\rightarrow - \chi_{R,\alpha}$, which also prohibits any mixing
between left and right moving modes. Without interactions,
the classification of this topological superconductor is obviously
$\mathbb{Z}$. Even including interactions that preserve this
$Z_2$ symmetry, the edge state in Eq.~\ref{1d} with four channels
of nonchiral Majorana fermions is still topologically stable,
namely it cannot be gapped out without breaking the $Z_2$
symmetry~\cite{fidkowski1,fidkowski2,zhangz8,qiz8,yaoz8,levinguz8}.



In this work we considered electronically driven superconductivity in
graphene moir\'e superlattices within a minimal single band triangular
lattice Hubbard
model with spin+orbital degeneracy.  We found that the valley degree of
freedom of graphene has qualitative effects on the superconductivity
compared to single-orbital Hubbard systems, favoring topological
$d+id$ paired spin-triplet states.  With SU(2) spin-rotation symmetry,
these states support exotic charge $4e$ pairing and half-vortices at
non-zero temperature.  Thus graphene may become not only a venue for
strong correlation physics, but also topological superconductivity.

Subsequent to the posting of the first version of this preprint, a
number of papers appeared emphasizing the physics related to Dirac
band crossings between a {\em pair} of flat mini-bands appearing
in some models of the moir\'e superlattice\cite{po2018origin,yuan2018model}. In this situation the
two low energy bands are intertwined and there is no obvious
separation between them. Consequently, in the strong coupling
limit, the minimal description is  two band
honeycomb lattice model, for which both the kinetic energy and
interactions are more complicated than in Eq.~\eqref{h1}.  In
the intermediate correlation regime, states near the Fermi energy
dominate the superconductivity, and it is not clear that either the
additional band far from the Fermi energy or any symmetry protected
Dirac point plays a major role.  
The major difference between Ref.\cite{po2018origin} and our own work is that the
former invokes ($SU(4)$) ferromagnetism, while we rely upon the
effective anti-ferromagnetic interaction, Eq.~\eqref{hj}.  Na\"ive
arguments beginning from the former perspective\cite{po2018origin}
appear to favor more conventional singlet superconductivity, which
need not be topological.  However, ferromagnetism in Hubbard models is
notoriously hard to find, and anti-ferromagnetism may be more robust
even when the honeycomb description is more appropriate.  The
compelling simplicity of the triangular framework suggests that
graphene moir\'e heterostructures which realize the single
band triangular regime are favorable for realizing topological
physics. This constitutes a design goal which is realizable
theoretically and experimentally.

Further studies should address these states quantitatively, the
possibility of quantum spin liquid physics in the Mott states, and
the effects of perturbations to the minimal Hubbard descriptions such
as disorder, magnetic fields, and more.  The low energy scale of these
graphene superlattices allows vastly larger tuning of doping and
magnetic field axes in comparison to conventional correlated
transition metal compounds, and their pure two-dimensionality makes
probing strictly Zeeman effects also possible.  Our results may serve
as guidance for such future studies.

\begin{acknowledgments}
LB thanks Andrea Young for explanations of flat band physics in
graphene moir\'e superlattices, and Lucile Savary for discussions of
group theory, pairing and Fierz identities in multi-orbital
systems. The authors are supported by the NSF materials theory
program through grant DMR1506119 (LB), and the David and Lucile
Packard Foundation and NSF Grant No. DMR-1151208 (CX).
\end{acknowledgments}
\bibliography{MAG}

\end{document}